 \documentstyle[prl,aps,multicol,amstex]{revtex}
\input{epsf}

\DeclareMathSymbol{\square}       {\mathord}{AMSa}{"03}
\DeclareMathSymbol{\blacksquare}  {\mathord}{AMSa}{"04}
\DeclareMathSymbol{\lozenge}      {\mathord}{AMSa}{"06}
\DeclareMathSymbol{\blacklozenge} {\mathord}{AMSa}{"07}
\DeclareMathSymbol{\vartriangleright}{\mathrel}{AMSa}{"42}
\DeclareMathSymbol{\vartriangleleft} {\mathrel}{AMSa}{"43}
\DeclareMathSymbol{\blacktriangledown}  {\mathord}{AMSa}{"48}
\DeclareMathSymbol{\blacktriangleright} {\mathrel}{AMSa}{"49}
\DeclareMathSymbol{\blacktriangleleft}  {\mathrel}{AMSa}{"4A}
\DeclareMathSymbol{\vartriangle}        {\mathrel}{AMSa}{"4D}
\DeclareMathSymbol{\blacktriangle}      {\mathord}{AMSa}{"4E}
\DeclareMathSymbol{\triangledown}       {\mathord}{AMSa}{"4F}

\begin{document}

\draft
\title{Bistability of Slow and Fast Traveling Waves in Fluid Mixtures}
\author{St.~Hollinger, P.~B\"uchel, and M.~L\"ucke}
\address{Institut f\"ur Theoretische Physik, Universit\"at des Saarlandes,
         Postfach 151150, D 66041 Saarbr\"ucken, Germany}
\maketitle

\begin{abstract}
The appearence of a new type of fast nonlinear traveling wave
states in binary fluid convection with increasing Soret effect
is elucidated and the parameter range of their bistability with 
the common slower ones is evaluated numerically. The 
bifurcation behavior and the significantly different
spatiotemporal properties of the different wave states - e.g. frequency,
flow structure, and concentration distribution - are determined
and related to each other and to a convenient measure of
their nonlinearity. This allows to derive a limit for the
applicability of small amplitude expansions. Additionally an universal 
scaling behavior of frequencies and mixing properties is found.
\end{abstract}
%
% 47.20.-k : Hydrodynamic stability
% 47.10.+g : General theory (of fluid dynamics)
% 47.20.Ky : Nonlinearity (including bifurcation theory)
%
%****************************************************************
%*                                                              *
%*      begin of text                                           *
%*                                                              *
%****************************************************************
%\narrowtext

\begin{multicols}{2}
In binary fluid mixtures heated from below there is an interesting 
feed-back loop between the fields of concentration, velocity, and 
temperature: The buoyancy force that drives the convective flow is 
changed by concentration variations. They in turn are 
{\it produced} via the Soret effect \cite{LL66,PL84} by temperature
gradients, i.e., via thermodiffusion and {\it reduced} by dissipative
concentration diffusion and by mixing through the convective flow. 
This coupling chain causes a surprising richness of spatiotemporal 
pattern formation \cite{CH93} even close to the onset
of convection. In particular there are convective structures 
\cite{PL84,CH93,WKPS85,TPC96,AR86,GB86,MS88,OYSK90,EOYSBLKK91,ZM91,%% 
WK92,lPES96,BLHK89,BLKS95I}
consisting of coupled traveling waves (TWs) of velocity, temperature, 
and concentration with significantly different shapes 
\cite{BLHK89,BLKS95I}. Since nonlinear concentration
advection is typically much larger than linear diffusion --- the 
ratio of these two transport rates can easily be above 1000 --- it 
is not surprising that these TW states are typically strongly nonlinear. 

In this letter we elucidate how with increasing Soret coupling 
there appear two different bistable TWs --- one about twice as fast 
as the other --- which both stably coexist with the stable quiescent 
conductive state. The convective amplitude of the
fast (slow) TW is small (large) while the amplitude of its 
concentration contrast is large (small). The fast stable TWs have 
so far remained unnoticed in experiments
\cite{WKPS85,TPC96,AR86,GB86,MS88,OYSK90,EOYSBLKK91,ZM91,WK92,lPES96}
and numerical simulations \cite{BLHK89,BLKS95I}.
They develop with increasing Soret coupling via a saddle
node bifurcation out of a dent in the subcritically bifurcating unstable
TW branch. They are most easily accessible via a two-loop hysteresis at 
larger but experimentally realizable Soret effects. Furthermore, 
we discovered an universal scaling
behavior of TW frequencies and mixing properties. And we found that 
the ratio of flow and phase velocity is a relevant parameter and a convenient
measure of their nonlinearity that allows to determine where an amplitude 
expansion around the onset breaks down. 

To investigate roll like convection structures
in a horizontal layer of, say, ethanol-water with Lewis number $L=0.01$ and
Prandtl number $\sigma = 10$ 
we determine the convective solutions that bifurcate
with a lateral periodicity length $\lambda =2$ out of the 
quiescent heat conducting state \cite{lengthscales}.
A finite difference method \cite{MACGezerre}
as well as a many mode Galerkin scheme is used
to solve the appropriate field equations 
\cite{LL66,BLKS95I}
in a vertical cross section through the rolls perpendicular
to their axes. Horizontal boundaries at top  
and bottom, $z=\pm1/2$, are no slip, perfectly heat 
conducting, and impermeable.
Our control parameter $r=R/R^0_c$ measuring the thermal 
driving is the Rayleigh number $R$
reduced by the critical one $R^0_c=1707.762$ for onset of 
convection in a pure fluid.
We are interested here in 
negative separation ratios $\psi$ \cite{CH93}. Then the Soret coupling
between deviations $T$ of temperature and $C$ of concentration
from their means tends to increase (decrease) the ethanol 
concentration in the cold (warm) fluid regions.

\begin{figure}
\epsfxsize=.68\hsize
\centerline{\mbox{\epsffile{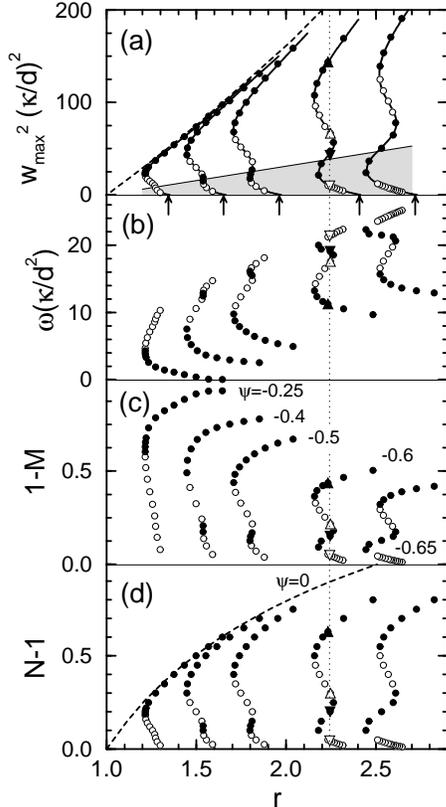}}}
\narrowtext
\caption[]{Evolution of TW--bifurcation diagrams with Soret coupling
           strength $\psi$: (a) squared maximal vertical flow $w_{max}^2$,
           (b) frequency $\omega$, (c) mixing number $1-M$, and (d)
           convective contribution to the Nusselt number $N-1$ {\it vs\/} reduced
           Rayleigh number $r$. Stable (unstable) TW states are marked by 
	   filled (open) symbols. Four of them on the vertical dotted line at $r=2.24$,
           $\psi=-0.6$ are identified for later discussion by 
           different triangles. Arrows mark Hopf thresholds $r_{osc}$ for
           onset of TW convection. The $\psi=0$ pure fluid limit is
           included in (a) and (d) by the dashed line. Full lines through
           the data points of (a) represent the fit discussed in the
           text. Only states in the shaded region of (a) are weakly
           nonlinear (cf.~text).}
\label{fig1}
\end{figure}

In Fig.~\ref{fig1} we show how an increasing Soret coupling strength 
changes the bifurcation diagrams of (a) maximal vertical flow 
intensity $w^2_{max}$, (b) TW frequency $\omega$, (c) mixing number 
$1-M$, and (d) convective contribution $N-1$ to the Nusselt number.
The order parameter 
$1-M=1-\sqrt{\langle C^2\rangle/\langle C^2_{cond}\rangle}$
measures the convective mixing. It is defined such as to vanish
in the conductive state, $C_{cond}= -\psi z$, and it approaches $+1$ for
 convection 
with strong mixing properties where the spatially averaged square of 
the concentration variation $<C^2>$ becomes small.
The overall subcritical bifurcation topology is caused by 
the interplay of two adverse effects:
{\it (i)\/} The Soret coupling to the degrees of freedom of the 
concentration field stabilizes the quiescent basic state
 \cite{HJ71},
since the Soret induced concentration distribution
reduces the buoyancy force that drives convection.
This shifts with increasing $|\psi|$ the TW bifurcation threshold
$r_{osc}$ (arrows in Fig.~\ref{fig1}) upwards along the r axis. 
{\it (ii)\/} With increasing convection mixing 
advectively reduces the Soret induced
concentration gradients and with it the influence of 
the Soret effect on the buoyancy so that the convection 
behavior of the mixture approaches that of the pure 
fluid -- dashed curves labelled
$\psi=0$ in Fig.~\ref{fig1}a, d.

\begin{figure}
\epsfxsize=0.9\hsize
\centerline{\mbox{\epsffile{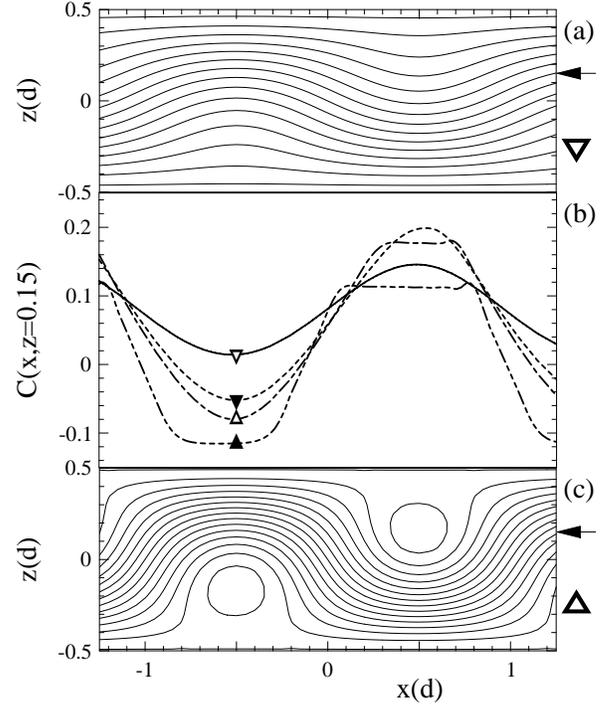}}}
\narrowtext
\caption[]{(b) Wave profiles of concentration deviation $C$
           (scaled by $\Delta T\alpha/\beta$) at the vertical
           position marked by arrows in (a) und (c). Therein streamlines in
           the frame comoving with the TW are shown. Symbols identify
           states in Fig.~\ref{fig1} on the dotted vertical line.}
\label{fig2}
\end{figure}

We found that the spatiotemporal properties of TW
states substantially change when the convective flow $w_{max}$ 
becomes larger than the phase velocity $v=\omega \frac{\lambda}{2\pi}$.
Starting
\noindent
at $r_{osc}$ with a large Hopf frequency $\omega_H$ and $w_{max}=0$ 
the frequency (Fig.~\ref{fig1}b) monotonically decreases 
along the TW solution branch while the convection amplitude $w_{max}$ grows.
States with $\chi=w_{max}/v<1$ (shaded region 
in Fig.~\ref{fig1}a) are weakly nonlinear,
while those with larger $\chi$ are strongly nonlinear.
In the former there are only open streamlines in the frame comoving with 
the TW (Fig.~\ref{fig2}a) and the concentration wave profile is basically
 harmonic
(Fig.~\ref{fig2}b $\triangledown$ and $\blacktriangledown$).
On the other hand, 
for $\chi>1$ there are also regions of closed streamlines
(Fig.~\ref{fig2}c). With growing $\chi$ their size and 
with it the anharmonicity of the concentration wave (Fig.~\ref{fig2}b,
$\vartriangle$ and  $\blacktriangle$) 
increases, since within regions 
of closed streamlines concentration is diffusively
mixed to 
alternatingly high and low plateau levels. When moving along a TW
bifurcation branch one observes with growing $\chi$ first an increase of the 
amplitude of a harmonic concentration wave up to $\chi \simeq 1$ and then
a decrease in amplitude combined with its wave profile
becoming more 
\noindent
and more trapezoidal (Fig.~\ref{fig2}b).

\begin{figure}
\epsfxsize=0.94\hsize
\centerline{\mbox{\epsffile{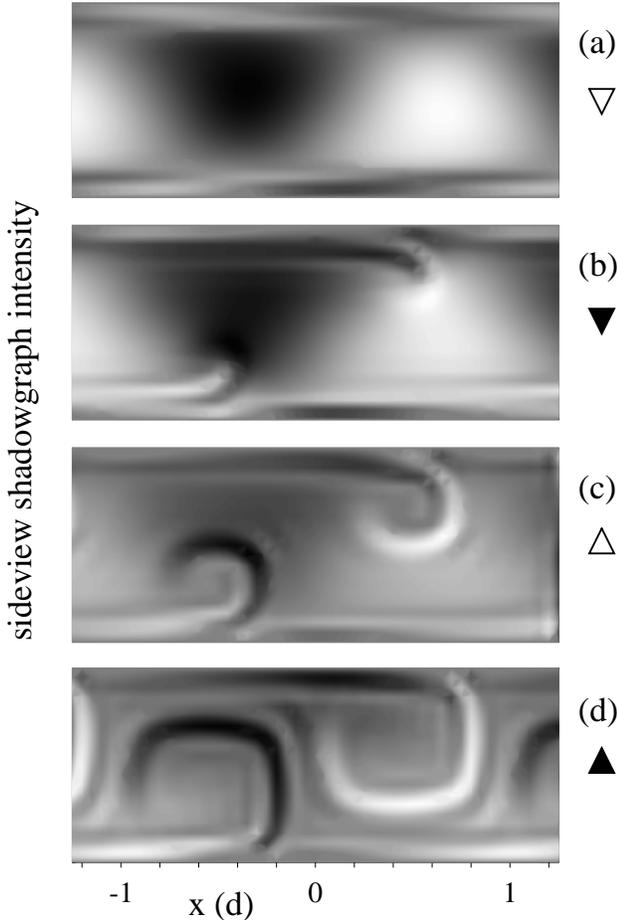}}}
\narrowtext
\caption[]{Sideview shadowgraph intensity distributions obtained
           as described in \cite{EOYSBLKK91,BLKS95I} from the
           numerically determined TW states that are marked by the
           respective triangles in Fig.~\ref{fig1}.}
\label{fig3}
\end{figure}

The structural changes that occur with growing $\chi$ can be 
observed experimentally in topview and even better in sideview shadowgraph 
intensity distributions \cite{KWM88,EOYSBLKK91,WK92}.
The sideviews in Fig.~\ref{fig3} of four TWs (marked by triangles 
in Fig.~\ref{fig1}) existing at the same $r \approx 2.24$ 
show in (a) the smooth
distribution of a weakly nonlinear linear
state ($\triangledown$) and in (b)-(d) strongly
nonlinear ones. In the latter the Soret 
induced ethanol rich (poor) boundary layers near the top (bottom) plate 
increasingly discharge via concentration jets into the
growing roll like regions
of closed  streamlines. Therein, the concentration is then homogenized
on alternating high and low levels.
%Thus the competing bistable TW states
%(e.g.,\ $\blacktriangledown$ and $\blacktriangle$) 
%cannot only be distinguished by their 
%different frequencies ($\omega_\blacktriangledown \simeq 2 
%\omega_\blacktriangle$)
%and Nusselt numbers but also 
%by their shadowgraphs.

\begin{figure}
\epsfxsize=0.95\hsize
\centerline{\mbox{\epsffile{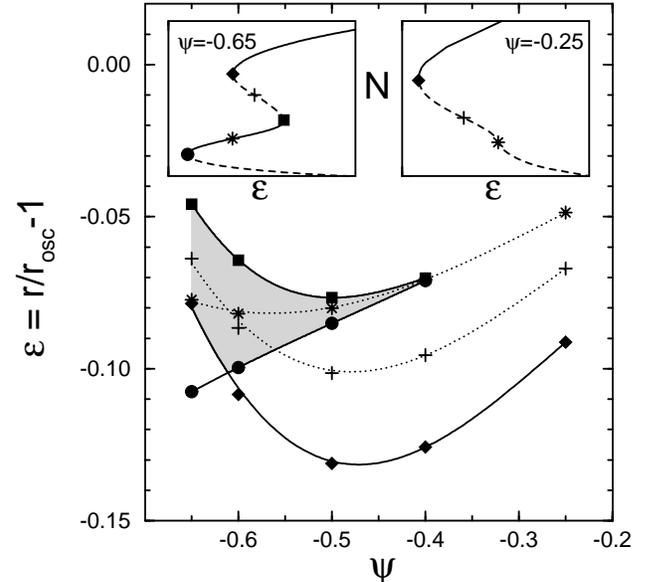}}}
\narrowtext
\caption[]{$\psi$--$\epsilon$ phase diagram of TW states at subcritical
           driving. In the shaded parameter regime two stable TWs exist.
           Bifurcation diagrams of $N$ {\it vs\/} $\epsilon$ in the insets
           explain the meaning of the symbols: $\mbox{\Large $\bullet$}$,
           $\blacksquare$, and $\blacklozenge$ represent saddle node states
           with small, medium, and large convection amplitude (large,
           medium, and small frequency), respectively. The
           first (second) inflection point is denoted by $\ast$ ($+$).}
\label{fig4}
\end{figure}

With increasing Soret coupling strength $|\psi|$ the TW solution branches 
in Fig.~\ref{fig1} become more and more contorted.
Already at $\psi\simeq -0.001$ the lower unstable branches of $w^2_{max}$ 
and $N$ have two inflection points with a dent in
between that can 
clearly be seen in Fig.~\ref{fig1} at $\psi=-0.25$.
The significance of such a dent on the lower unstable branch of solutions
being unaccessible to experiments 
\cite{WKPS85,TPC96,AR86,GB86,MS88,OYSK90,EOYSBLKK91,ZM91,WK92,lPES96}
as well as to earlier numerical
simulations \cite{BLHK89,BLKS95I} 
has not been appreciated appropriately although it can
be seen in the paper of Bensimon et al.~\cite{BPS89} in the limit
of small Soret coupling and $\sigma\rightarrow\infty$.
With our new finite difference method and our many mode Galerkin scheme
we can now trace out the complete solution branch and 
determine how its bifurcation topology changes with $\psi$. 
\noindent
We found that
the above mentioned dent develops at 
$\psi\simeq -0.4$ into a forwards bending arc when the first inflection
point ($\ast$ in Fig.~\ref{fig4}) emits two new saddles ($\blacksquare$ and 
$\mbox{\Large $\bullet$}$ in Fig.~\ref{fig4}). So in the shaded region of
the control 
parameter plane of Fig.~\ref{fig4} three stable states occur: For
$\psi\lesssim -0.4$ two different 
stable TWs coexist --- one fast, the other one slow --- that compete with
each other and with the stable 
quiescent basic state. Furthermore, a new scenario appears 
for $\psi \lesssim -0.61$. There, the Rayleigh
number $r^S_{\mbox{\large $\bullet$}}$ of the saddle
with small $N,w_{max}$ and large $\omega$ drops below the 
saddle $r^S_{\mbox{$\blacklozenge$}}$ with large $N,w_{max}$ and small $\omega$ so that a 
two-step hysteresis opens up between basic state, slow TWs, and fast
TWs \cite{endwalls}.

The first inflection point ($\ast$ in Fig.~\ref{fig4}) of the TW~bifurcation 
branch coincides closely with the 
boundary $\chi=1$ between weakly and strongly nonlinear states --- the change
in curvature of the bifurcation branch
reflects the
change in the TW structure.
Even for $\psi$ as small as about $-0.001$
amplitude equations of quintic order might
possibly reproduce realistically only those TW states that
lie between $r_{osc}$ and the first inflection point.
The fact that
$\chi\simeq1$ is an upper limit to amplitude expansions around
\noindent
the Hopf bifurcation threshold follows also directly from the
behavior of the multivalued functional dependence of the flow intensity
$X=w_{max}^2$ on $r$. Its singlevalued inverse $r(X)$
--- to see it just turn Fig.~\ref{fig1}a by $90^o$ --- can be
represented extremely well by the product
\begin{equation}
\label{rationalfit1}
r(X) = r_0(X)\,Q(X)
\end{equation}
cf.~the full lines in Fig.~\ref{fig1}a. Here $r_0(X)$ is the $\psi=0$
border curve for the pure fluid (dashed line in Fig.~\ref{fig1}a) that
starts at $r_0(X=0)=1$ and grows for large $X$ slightly sublinearly. The
rational function
$$
Q(X) = \frac{a_0+a_1X+a_2X^2+a_3X^3}{1+b_1X+b_2X^2+b_3X^3}
$$
ensures with $a_0=r_{osc}$ that $r(X=0)=r_{osc}$ and with finite
${a_3}/{b_3}$ that $r-r_0$ does not diverge for large $X$. 
With the third order polynomials in $Q$ four different $X$ can 
have the same $r$. In that sense (\ref{rationalfit1})
is a minimal representation of the
bifurcation diagrams of Fig.~\ref{fig1}a. Now, the radius of
convergence $X_c$ of an amplitude expansion of $Q$ in a power series 
in $X$ is
given by the absolute value of that pole of $Q$ closest to $X=0$ in the
complex $X$--plane.
Our fits for $Q(X)$ show that this pole lies on the
negative real  $X$--axis with $X_c$ 
close to $v^2$ so that $\chi=w_{max}/v \simeq 1$ is indeed an 
{\it upper limit\/} for the applicability of amplitude expansions.

A recent TW model \cite{HL96} into which enters $\chi$ as a
relevant parameter predicts that the frequencies are
universally determined
by the "distance" $r(X) - r_0 (X)$ of the 
\begin{figure}
\epsfxsize=0.8\hsize
\centerline{\mbox{\epsffile{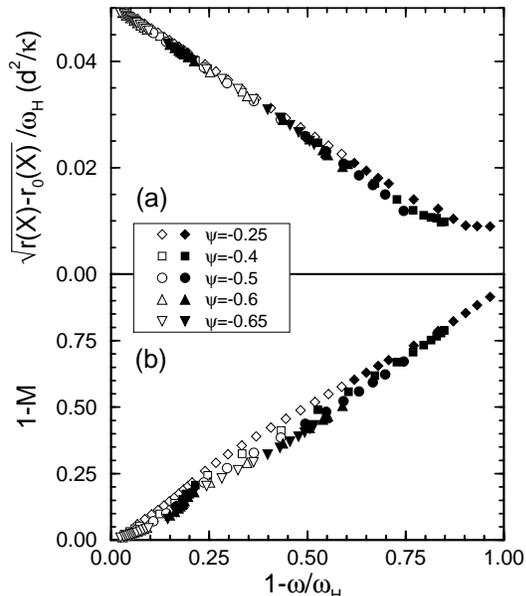}}}
\narrowtext
\caption[]{Universal, $\psi$ independent scaling relations between: (a)
           Frequency and "distance" $r-r_0$ of TW states from $\psi=0$
           convection (cf.~text), (b) degree of convective mixing $1-M$ and
           TW frequency. Here $\omega_H$ is the Hopf frequency at onset. 
	   Shown are all TW states of Fig.~\ref{fig1} and some that have 
	   been suppressed there for clarity (stable TWs -- filled symbols,
           unstable TWs -- open symbols).}
\label{fig5}
\end{figure}
\noindent
TW states from the pure 
fluid convection state ($\psi = 0$).
 For any TW with a particular $X = w^{2}_{max}$ this
distance parallel to the $r$-axis can be read off in Fig.~\ref{fig1}a.
We confirm this prediction in Fig.~\ref{fig5}a. It shows that the
scaled frequencies $1 - \omega / \omega_{H}$ are indeed linearly related to
$\sqrt{r - r_0} / \omega_{H}$ in an
universal, $\psi$-independent manner that holds for stable as well as for
unstable TWs. Furthermore, according to Fig.~\ref{fig5}b the
convective mixing
$1 - M$ grows linearly with the reduced frequency deviation
$1 - \omega / \omega_{H}$ from the Hopf frequency and also
$1 - M$ is roughly linearly related to $\sqrt{r - r_0} / \omega_{H}$.

%****************************************************************
%*                                                              *
%*      Summary							*
%*                                                              *
%****************************************************************

To summarize, we have found bistability between slow
and fast TWs and we have elucidated how the latter appear
when the Soret coupling becomes sufficiently strong. Test runs with
smaller Lewis and/or larger Prandtl numbers lead to similar bifurcation
behavior. To check our results experimentally ---
preferably in annular containers --- and also to investigate effects
related to spatially confined states, we suggest to use mixtures with
$\psi\simeq-0.6$ as in Refs.~\cite{WKPS85,TPC96}.

%****************************************************************
%*                                                              *
%*      Acknowledgements					*
%*                                                              *
%****************************************************************
{\it Acknowledgements\/} --- This work was supported by the  Deutsche
Forschungsgemeinschaft. A graduate scholarship of the Saarland
for one of us (SH) is gratefully acknowledged.

%****************************************************************
%*                                                              *
%*      list of references                                      *
%*                                                              *
%****************************************************************

\end{multicols}


\begin{references}
\bibitem{lengthscales}
We scale lengths by the thickness $d$ of
the fluid layer,
times by the vertical diffusion time $d^2/\kappa$, deviations $T$ 
from the mean temperature by the imposed vertical 
temperature difference $\Delta T$, and deviations $C$ from the mean
concentration  by $\Delta T \alpha/ \beta$
with $\alpha$ ($\beta$) being the thermal (solutal) expansion 
coefficient.
$L=D/ \kappa$ is the  Lewis number  and $\sigma=\nu/ \kappa$ the Prandtl
number. Here $\nu$ is the 
kinematic viscosity, $\kappa$ the thermal diffusivity, and
D the concentration diffusion coefficient.
\bibitem{MACGezerre}  
We use a finite difference algorithm \cite{BLKS95I} that is modified by a 
multigrid method for pressure iteration and a DuFort-Frankel scheme 
to solve the momentum balance equations \cite{Jung}.
The unstable TW solutions are determined with a linear control process in
which the temperatures at the plates are varied.
\end{references}
\end{document}